\documentclass[twocolumn,twoside,slac_two]{revtex4}
\usepackage{graphicx}
\usepackage{fancyhdr}
\begin{document}

%Title of paper
\title{{\it Fermi} and {\it Swift} observations of the bright short
  GRB 090510}

\author{V. Pelassa}
\affiliation{LPTA, CNRS/IN2P3 - Universit\'e Montpellier 2}
\author{M. Ohno}
\affiliation{ISAS/JAXA}
\author{on behalf of the {\it Fermi} LAT and GBM collaborations}
\begin{abstract}
  The bright short-hard GRB 090510 was observed by both {\it Swift} and
  {\it Fermi} telescopes. The study of the prompt emission by Fermi
  revealed an additional high-energy spectral component, the
  largest lower limit ever on the bulk Lorentz factor in a short
  GRB jet, and brought the most stringent constraint ever on
  linear Lorentz invariance violation models. The fast repoint and
  follow-up by both telescopes allowed the first multiwavelength
  study of a GRB afterglow from optical range to several
  GeV. This long-lived emission has been studied in the framework of the internal
  shock and external shock models.
\end{abstract}

\maketitle

\thispagestyle{fancy}

\section{OBSERVATIONS}

On May 10th, 2009 at 00:23:00.48 UT, the Burst Alert Telescope (BAT)
onboard the {\it Swift} observatory detected a bright Gamma-Ray Burst
(GRB) \cite{GCN9331}, and autonomously 
repointed towards it within 100 s, allowing for over 2 days of
follow-up observation by the Ultra-Violet and Optical Telescope (UVOT) and
the X-Ray Telescope (XRT). An afterglow was detected in ultra-violet and
X-ray, which analysis provided a location
RA = 333.55227, Dec = -26.5827, with an error of 1.4 arcsec (90\%
statistical) \cite{GCN9339}. VLT afterglow observations
provided a spectroscopic redshift z = 0.903 $\pm$ 0.003
\cite{GCN9353}.

\begin{figure}
\begin{center}
\includegraphics[width=0.99\linewidth,angle=0]{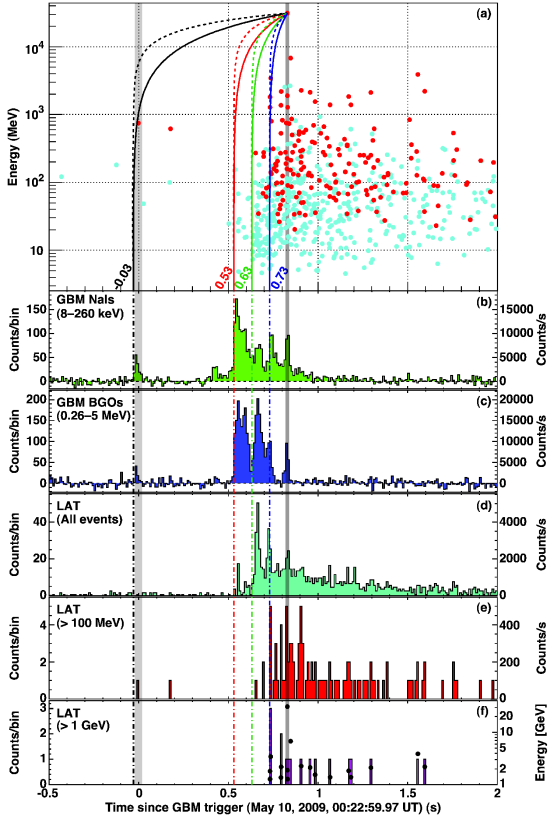}
\end{center}
\caption{GRB 090510 observation by {\it Fermi} \cite{Nat090510}.\\
  {\bf Panel (a) :} LAT events passing the onboard
  (blue) or on-ground (red) photon selections. 
  The lines show the arrival time dependence on photon energy for the
  events associated to the 31 GeV photon, according to linear (solid
  lines) or quadratic (dashed lines) energy dependence in the LIV
  model (see section~\ref{sect23}). Different parts of the
  lower-energy emission are considered : full GBM emission (black),
  bulk GBM emission (red), bulk LAT emission (green), coicident GBM
  peak (blue).\\ 
  {\bf Panels (b -- f) :} multi-instrument lightcurve. Curves (b -- d)
  are background-subtracted. Curve (d) 
  shows all LAT events passing the onboard photon selection. Curves (e -- f)
  show on-ground selected photons inside the region of
  interest, used for the spectral analysis. Panel (f) also shows event
  energies.}
\label{fig:lc_prompt}
\end{figure}

On the same day at 00:22:59.97 UT, the Gamma-ray Burst Monitor
(GBM) onboard {\it Fermi} detected a
bright GRB \cite{GCN9336}. A small and short spike
at trigger time was followed 0.5 s later by 0.5 s-long emission made
of 7 pulses.
This GRB also triggered the {\it Fermi} Large Area Telescope (LAT) onboard
algorithm \cite{GCN9334} which located it at RA = 333.4$^\circ$, Dec =
-26.767$^\circ$, with an error of 7 arcmin (statistical), consistent with the
GBM and {\it Swift} detection. The spacecraft autonomously repointed
towards the source for 5 hours of follow-up observation. The prompt
emission was coincident with the GBM main emission. More than 150
photons above 100 MeV were observed, and more than 20 events above 1
GeV in the first minute after trigger (see fig.~\ref{fig:lc_prompt})
\cite{GCN9350}. The follow-up observation allowed the detection of a
significant GeV emission up to 200 s after trigger.

Also AGILE \cite{GCN9343}, Integral-SPI and Suzaku-WAM \cite{GCN9355} detected
GRB 090510 prompt emission. According to all observations GRB 090510
was a very bright short GRB (table~\ref{tab:t90_t50}).

In the following we focus on the prompt emission seen by {\it Fermi}
(section~\ref{sect2}) and the multiwavelength afterglow study based on
{\it Swift} and {\it Fermi} data (section~\ref{sect3}). {\bf
  Convention :} spectral and temporal indices are defined according to
to the fluence evolution description : $F\;\propto\;
t^{-\alpha}\,\nu^{-\beta}$

% tableau T90, T50
\begin{table}
\caption{Duration ($T_{90}$ and $T_{50}$) of GRB 090510
  prompt emission observed by different instruments. \cite{ApJ090510}}
\label{tab:t90_t50}
\begin{center}
\begin{tabular}{|c|c|c|}
\hline
Instrument & $T_{90}\;(s)$ & $T_{50}\;(s)$ \\
\hline
GBM/NaI 3,6,7 & 2.1 & 0.2 \\
{\it Swift}/BAT & 4.0 & 0.7 \\
Integral-SPI & 2.5 & 0.1 \\
Suzaku-WAM & 5.8 & 0.5 \\
\hline
\end{tabular}
\end{center}
\end{table}

% courbe de lumière émission prompte

\section{THE $\gamma$-RAY PROMPT EMISSION}
\label{sect2}

\subsection{Spectroscopy}

\begin{figure}
\begin{center}
\includegraphics[width=0.99\linewidth,angle=0]{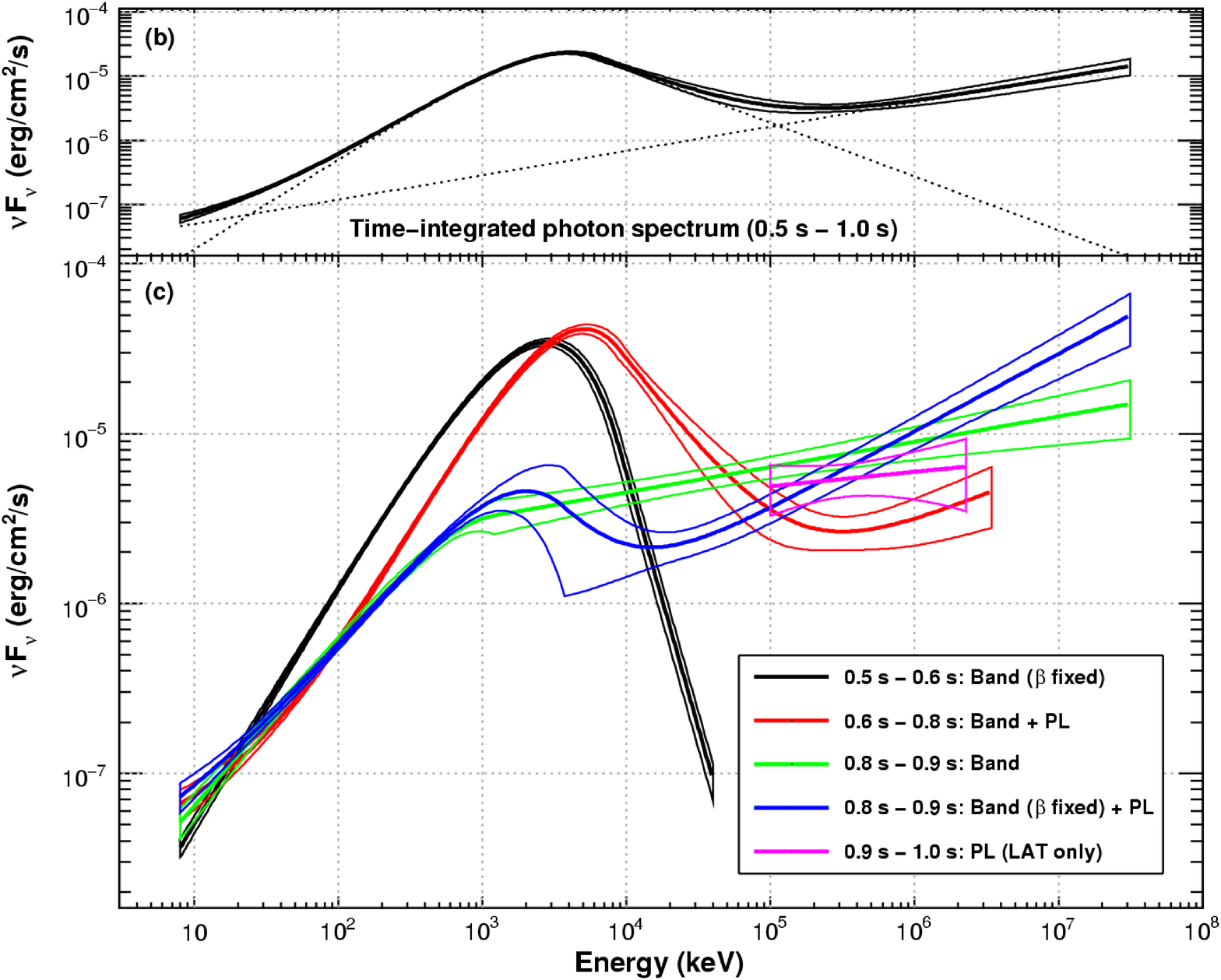}
\end{center}
\caption{GRB 090510 prompt spectrum \cite{ApJ090510}. \\
  {\bf Panel (b) :} time-integrated Spectral
  Energy Distribution (SED). The Band function and power-law
  components are shown. Their sum is drawn in thick line with 68\%
  confidence level contours. {\bf Panel (c) :} time-resolved SED with
  68\% confidence level contours. In bin 'a' no
  signal is detected in the LAT. Bins 'b' and 'c' show an additional
  power-law component. No signal is detected in the GBM after 0.9 s
  (bin 'd').}
\label{fig:sed_prompt}
\end{figure}

The spectral analysis of the prompt emission was performed with
RMfit, combining GBM and LAT data (fig.~\ref{fig:sed_prompt}).

The time-integrated spectrum shows a significant additional high-energy
power-law component ($N_\sigma = 5.6$, $\beta_{PL} = 0.62 \pm 0.03$),
which carries 37\% of the total $\gamma$-ray fluence.

The time-resolved spectroscopy shows the late arrival of the LAT
emission and of the additional component. In bin 'a' (0.5s -- 0.6s after
{\it Fermi} trigger), a Band function with steep $\beta_{Band} >$ 4 is
fitted. In bin 'b' (0.6s -- 0.8s), a significant additional power-law
component is found  ($\beta_{PL} = 0.66 \pm 0.04$). In bin 'c' (0.8s
-- 0.9s) this high-energy component can be fitted but not
significantly. In bin 'd' (0.9s -- 1.0s) the LAT data only are fitted
by a power-law ($\beta_{PL} = 0.9 \pm 0.2$).

\subsection{Limits on the jet's initial bulk Lorentz factor $\Gamma_0$}
\label{sect24}

The observation of high-energy events allows to 
constrain the jet's initial bulk Lorentz factor $\Gamma_0$.
Low- and high-energy photons emitted in the same physical region of
the jet can interact to produce $e^\pm$ pairs. The opacity to pair-production
is strongly reduced if the emission region moves towards
the observer with a high bulk Lorentz factor. The optical depth to
$e^\pm$-pair production at a given photon energy
$\tau_{\gamma\gamma}(E)$ depends on the $\gamma$-ray emission spectral
shape and variability time scale $t_v$, the GRB redshift and the jet's
bulk Lorentz factor. The condition $\tau_{\gamma\gamma}(E < E_{max}) <
1$, where $E_{max}$ is the maximal observed photon energy, yields a
lower limit on $\Gamma_0$.

For different epochs of the LAT emission, estimates of $t_v$
and spectral models lead to several possible lower limits, all of
order $\Gamma_{0,min} \sim 10^3$ (see 
table~\ref{tab:Gamma0}). This is the 
highest lower limit ever set on a GRB, and the
most rapid ejection ever observed from a short GRB.

% tableau limites Gamma0 papier ApJ
\begin{table}
\begin{center}
\caption{$\Gamma_{0,min}$ values from GRB 090510 observations.}
\label{tab:Gamma0}
\begin{tabular}{|c|c|c|c|c|}
\hline
Time bin & Spectral model & $t_v$ (ms) & $E_{max}$ (GeV) &
$\Gamma_{0,min}$ \\
\hline
b & Band + PL & 14 $\pm$ 2 & 3.43 & 951 $\pm$ 38 \\
d & Band & 12 $\pm$ 2 & 30.5 & 1324 $\pm$ 50 \\
d & Band + PL & 12 $\pm$ 2 & 30.5 & 1218 $\pm$ 61 \\   
\hline
\end{tabular}
\end{center}
\end{table}

\subsection{Interpretation of the high-energy spectrum}

The physical interpretation of the high-energy prompt spectrum of GRB
090510 is described in detail in \cite{ApJ090510}. The lower-energy
part of the prompt emission spectrum ($<$ 10 MeV) can be 
explained by non-thermal synchrotron processes, as well as
photospheric thermal emission.

In a non-thermal leptonic model framework, Synchrotron Self-Compton (SSC)
emission is expected at GeV-TeV energies and can result in the observed
additional power-law component. 
Detailed simulations of SSC emission have been performed, using the
observed Band component as the seed population, taking into account
internal $\gamma\gamma$ opacity, self-absorption, radiative escapes,
and considering either Thomson or Klein-Nishina scattering regimes. A
low magnetic field ($B<<10^{3}G$) favors Compton scattering and
can explain the observed bright and hard high-energy
component. It would also slow down the cascade formation which could
explain the late onset of the high-energy emission.
The interpretation of the high-energy emission as forward shock
emission from an early afterglow was also considered to explain
the delayed onset of the high-energy emission. However, this scenario
assumes a very rapid ejection ($\Gamma \simeq$ 2000 -- 4000), and
implies a high-energy component spectral index of $\beta_{PL} \simeq
1$. This spectrum is consistent with the long-lived spectrum, but not
with the additional component found in bin 'b' ($\beta_{PL} = 0.66$).
In both cases, the hard spectral index of the soft photons
($\alpha_{Band} = -0.48$) is not explained by the standard 
synchrotron mechanism.

In hadronic models, secondary $e^\pm$ pairs come from the decay of
pions formed through photohadronic processes and by the
$\gamma-\gamma$ attenuation of synchrotron photons radiated by protons
and ions. Synchrotron radiation or inverse Compton scattering from
these pairs can produce the observed high-energy $\gamma$-ray emission. 
In the case of photohadronic models, a stronger magnetic field makes
secondary pion production more efficient through more rapid hadronic
acceleration, but results in a high-energy component spectral index
$\beta_{PL} \simeq 1$, significantly softer than observed. 
On the other hand, inverse Compton scattering of secondary pairs can produce a
high-energy component as hard as the one observed, but this
scenario requires a very high isotropic-equivalent power $P_{iso} \sim
10^{55} erg.s^{-1}$. Proton synchrotron models also require
a high isotropic equivalent energy, and a high magnetic
field. However, the apparent $E_{iso}$ problem can be solved if the
jet is strongly collimated or if the true Lorentz factor $\Gamma$ is a
factor $\simeq$2 less than the minimum values $\Gamma_{0,min}$
estimated through the $\gamma-\gamma$ opacity constraint. If
$\Gamma>\Gamma_{0,min}$, then a proton synchrotron model is strongly
disfavored, though note that GRBs with a very high isotropic energy
have been observed.

% SED emission prompte

\subsection{Limits on Lorentz Invariance Violation (LIV)}
\label{sect23}

Some quantum gravity models allow for a dependence of the photons
speed $v_{ph}$ on their energy, i.e. their arrival time can be written
as a development in their energy. In this framework, a low-energy ($E_\ell$) and a
high-energy ($E_h$) photons emitted together would arrive at different
times, with the dominant LIV delay term :
\begin{eqnarray}
\Delta t & = & s_n \frac{(1+n)}{2 H_0} \frac{(E_h^n -
  E_\ell^n)}{(M_{QG,n}c^2)^n} \nonumber \\
 & \times & \int_0^z \frac{(1+z')^n}{\sqrt{\Omega_m
  (1+z')^3 + \Omega_\Lambda}} dz' \nonumber
\end{eqnarray}

where $z$ is the GRB redshift, and n = 1 or 2. A standard cosmology $(h,\,
\Omega_m,\, \Omega_\Lambda) = (0.71,\, 0.27,\, 0.73)$ has been assumed
for the computations. The difference in arrival times can be of any sign
$s_n$, depending on the models. The observation of high-energy photons
from distant sources allows to put constraints on the energy scales
$M_{QG,n}$ that appear in the development.

GRB 090510 has a redshift of $z = 0.903 \pm 0.03$, and its emission in
the LAT energy range includes a photon of energy $E_h = 30.5^{+
  5.8}_{- 2.6}$ GeV detected 0.829 s after the GBM
trigger. This event's topology makes it a very good photon candidate.
The expected background rate and this event's direction
proximity to GRB 090510 location allowed us to
associate this event to the burst with high confidence ($N_\sigma =
$4.4 to 5.6 depending on the selections used).

The observed lightcurve allows different measurements of the
temporal delay for this event, depending on which part of the lower
energy emission 
it is associated to. As described in detail in \cite{Nat090510},
several lower limits on the linear 
term's energy scale $M_{QG,1}$ could be derived (see
table~\ref{tab:limit_MQG}). All require $M_{QG,1} > M_{Planck}$, which
is a strong constraint on linear LIV models.

Lower limits on $M_{QG,2}$ were also derived, but they are much less
constraining. For more details on LIV limits derivation, see
\cite{Vlasios}.

% tableau limites M_QG
\begin{table}
\begin{center}
\caption{Limits on linear LIV derived from GRB
  090510 observations \cite{Nat090510}.}
\label{tab:limit_MQG}
\small
\begin{tabular}{|c|c|c|c|}
\hline
$|\Delta t/\Delta E|$ or $|\Delta t|$ & $M_{QG,1}/M_{Planck}$ & $s_n$
& Method \\
\hline
$<$ 30 ms/GeV & $>$ 1.22 & $\pm$ 1 & LAT lag
analysis \\
$<$ 859 ms & $>$ 1.19 & 1 & not emitted ~~\\
 & & & before GBM onset\\
$<$ 10 ms & $>$ 102 & $\pm$ 1 & GBM pulse
width \\
\hline
\end{tabular}
\normalsize
\end{center}
\end{table}

\section{AFTERGLOW MULTI-WAVELENGTH STUDY}
\label{sect3}

\subsection{eV to GeV observations}

% cb lumiere multi-lambda
\begin{figure}
\begin{center}
\includegraphics[width=\linewidth,angle=0]{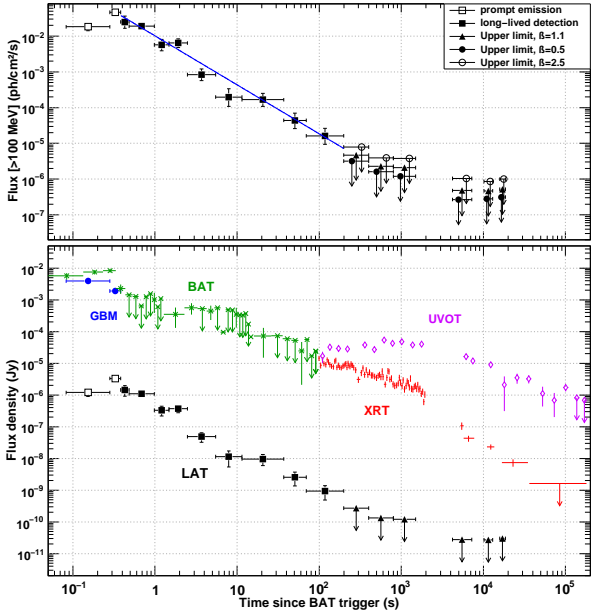}
\end{center}
\caption{ GRB 090510 {\it Swift} and {\it Fermi} observations
  \cite{MW090510}.\\
  {\bf Top panel :} LAT flux above 100 MeV and best fit to the flux
  decay (line). {\bf Bottom panel :} energy flux densities averaged in the
  observed energy bands: BAT (15 keV -­ 350 keV, stars); XRT (0.2 keV -­
  10 keV, crosses); UVOT renormalised to white (diamonds); LAT (100
  MeV -­ 4 GeV, filled squares; the average spectral index was used to
  convert from photon to energy flux) with upper limits for $\beta$ = 1.1
  (triangles). The prompt emission is shown for comparison: GBM (8 keV
  ­- 1 MeV, circles), LAT (100 MeV ­ 4 GeV, empty squares). XRT
  lightcurve is obtained as in \cite{Evans07,Evans09}. All other data are
  shown with 68\% error bars or 95\% confidence level upper limits.}
\label{fig:lc_afterglow}
\end{figure}

The fast repoint by {\it Swift} and the ARR performed by {\it Fermi}
provided a long-lasting observation of GRB 090510 afterglow, from
the optical range to several GeV, interrupted by Earth occultation of
the GRB to both observatories. The fluxes observed by all instruments
are reported in fig.~\ref{fig:lc_afterglow}.

UVOT and XRT observations of the afterglow started $\sim$ 100 s after
the prompt emission trigger, and lasted 
over 2 days. The optical and UV lightcurve is best fit by a smoothly
broken power-law with an initial rise
: $\alpha_{Opt,1} = -0.50^{+0.11}_{-0.13}$. The smooth break
at $t_{Opt} = 1.58^{+0.46}_{-0.37}$ ks is followed by a shallow decay
$\alpha_{Opt,2} = 1.13^{+0.11}_{-0.10}$. The X-ray lightcurve is best
fit by a broken power-law, with an initial shallow decay $\alpha_{X,1}
= 0.74 \pm 0.03$, a break at $t_X = 1.43^{+0.09}_{-0.15}$ ks and a
steep late decay $\alpha_{X,2} = 2.18 \pm 0.10$.

The beginning of the temporally-extended emission in the LAT has been
chosen as the end of the prompt emission seen by the GBM (0.9 s after
GBM trigger, i.e. 0.38 s after BAT trigger). A significant emission
was detected up to 200s after trigger and up to 4 GeV. This observation
was divided in time bins, in all of them an unbinned likelihood
spectral analysis was performed using diffuse class events
\cite{ApJ080825C}. The flux 
decay is best fit by a simple power-law of index $\alpha_\gamma = 1.38
\pm 0.07$, with no significant feature or break. The power-law
spectrum shows no significant evolution over time and has an average
index $\beta_\gamma = 1.1 \pm 0.1$.

\subsection{Discussion on the afterglow origin}

% 5 SED + SED at T+100s
\begin{figure}
\begin{center}
\includegraphics[width=0.99\linewidth,angle=0]{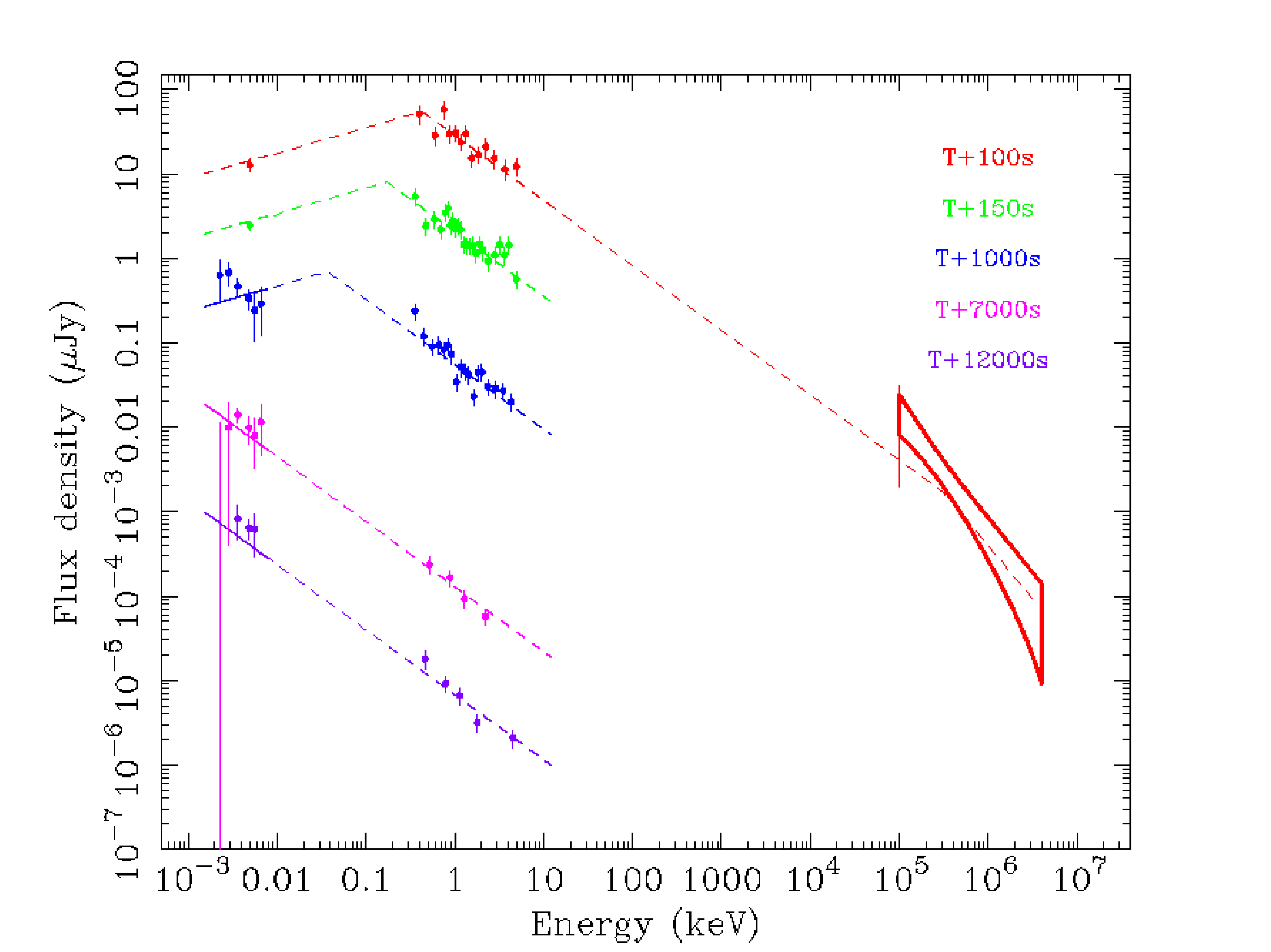}
\includegraphics[width=0.99\linewidth,angle=0]{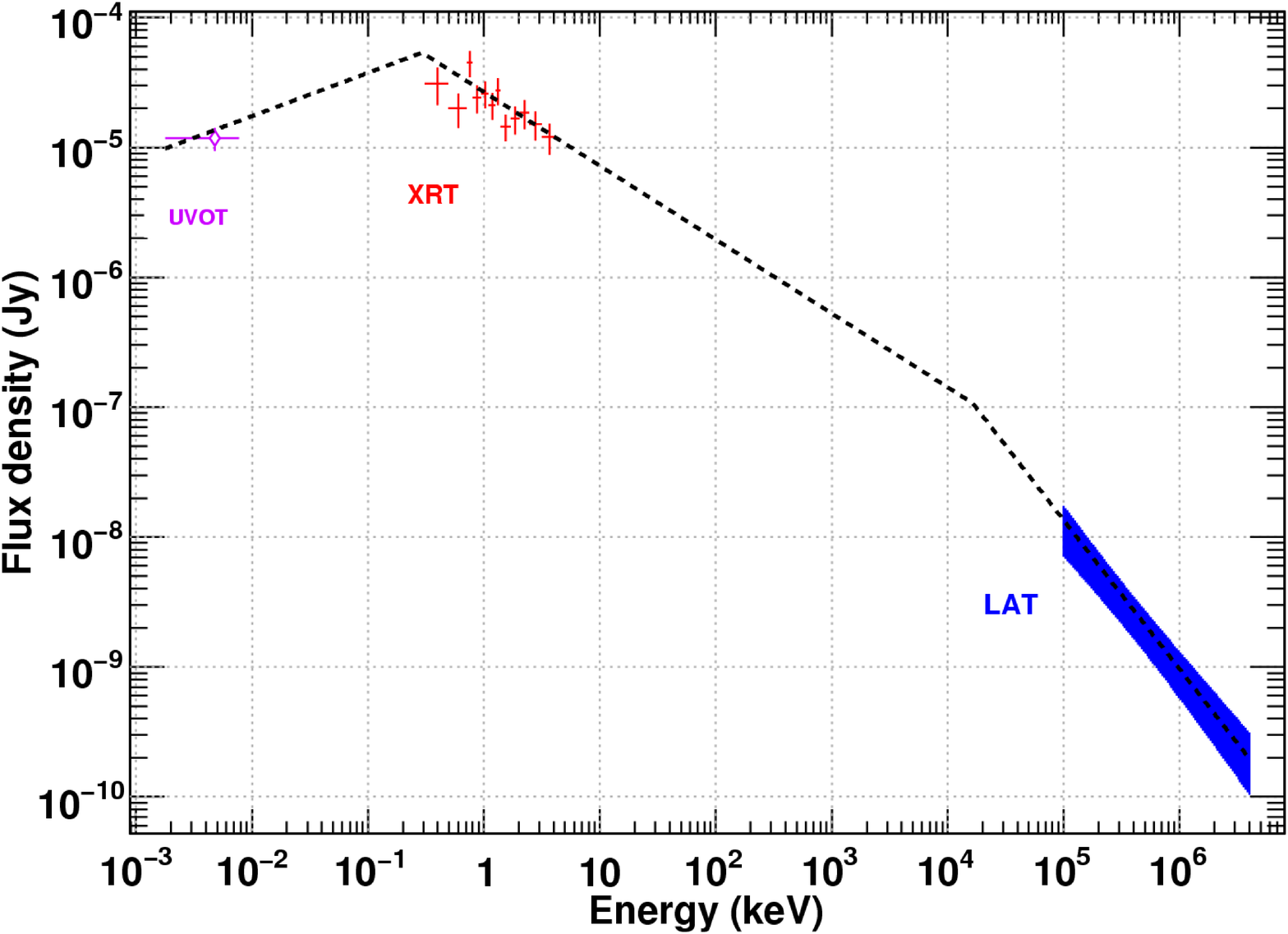}
\end{center}
\caption{GRB 090510 UVOT-XRT-LAT SEDs \cite{MW090510}.\\
  {\bf Top :} SEDs at different epochs, with the best fit shown
  ($\beta_2$ and $\beta_3$ constrained, see text). The
  butterfly at 100s indicates the 68\% confidence level region for the
  LAT flux, obtained from an unbinned likelihood analysis (95\% error
  bar at 100 MeV is shown). Successive SEDs in time order are rescaled
  by 1:1, 1:10, 1:100, 1:1000, 1:10000.
  {\bf Bottom :} SED at 100s, with the best fit shown. The $\beta_2$ -
  $\beta_3$ constraint is relaxed. 68\% error bars are shown for UVOT
  and XRT data. A 68\% confidence level region is drawn in the LAT
  energy range.}
\label{fig:sed_afterglow}
\end{figure}

The interpretation of the long-lasting multi-waveband emission fron
GRB 090510 is discussed in detail in \cite{MW090510}. Two scenarios
are considered in the fireball model frame.

In the first scenario, X-ray and $\gamma$-ray emissions are due
to internal shocks, and the optical and UV emission to the forward
shock. The fluxes measured at 100 s after trigger in the X-ray
and $\gamma$-ray ranges are consistent with an internal shock origin,
with some fine tuning required. For instance this scenario assumes
that the initial rise of the optical emission is due to the forward
shock onset, and therefore is steeper than what was observed. But 
the observation may have caught the end of the steep rise phase.
This interpretation
also requires a very low ambient density $n \sim 10^{-6}cm^{-3}$,
which is low, even for short GRBs. The assumed initial bulk
Lorentz factor $\Gamma_0 \sim 10^3$ is consistent with the prompt
emission analysis (see section~\ref{sect24}).

Another possibility is that the full long-lasting emission comes from the
forward shock region. This model predicts a broad spectrum with a 
doubly-broken power-law shape and constraints on the indices :
$\beta_1 = -1/3$, $\beta_3 = \beta_2 + 1/2$. Five successive Spectral
Energy Distributions (SED) have been built for 5 different observation
dates (100 s, 150 s, 700 s, 1000 s, 1400 s after trigger), including
X-ray and UV data, as well as LAT data for the SED at 150 s (see
fig.~\ref{fig:sed_afterglow}). A fit of 
these data yields a good agreement with the aforementioned model, with
$\beta_2 = 0.78 \pm 0.04$. The low-energy break decreases 
with time, from 0.43 keV to $<$0.01 keV. The high-energy break
is poorly constrained ($\in [10 - 130]$ MeV) but is confirmed by a fit
of the 100 s SED alone ($N_\sigma > 4.8 $) where the 
constraint on $\beta_2$ and $\beta_3$ is relaxed but still holds
within error bars. As a result, the afterglow emission spectrum is well
described by the forward shock model over 9 energy decades. The
lightcurves observed before 1 ks after trigger are also consistent with
this model. However, the early onset of the forward shock emission
requires a very high bulk Lorentz factor $\Gamma_0 > 5800$, and some
temporal emission properties are not well explained,
e.g. $\alpha_{X,1}$ is too shallow and $\alpha_{Opt,2} \neq
\alpha_{X,2}$. Theoretical extensions of the model can alleviate these
problems. E.g., a phase of energy injection 
or an evolution of the microphysical parameters of the blast wave may
cause an early shallow decay of the X-ray flux ; also the difference
between X-ray and optical late decay slopes could be explained by
dynamical effects.

\section{CONCLUSION}

GRB 090510 observation was remarkable for several reasons. First, it
was the first short GRB of known redshift ($z = 0.903$) observed at GeV
energies. It yielded the highest energy photon ever observed from a
short GRB (30.5 GeV). The observation of high-energy photons from this
GRB allowed to put strong and robust constraints on linear LIV models
($M_{QG,1}>M_{Planck}$ required) and on the jet's
velocity ($\Gamma_{0,min} \sim 10^3$).

This GRB also yielded the first clear evidence of an additional
high-energy spectral component ($N_\sigma = 5.6$ on the
time-integrated prompt spectrum), which was to be observed in several
other bursts \cite{SGuiriec,ApJ090902B}. This high-energy spectral
component is a hint for SSC radiation or a possible UHECRs production
in GRBs.

Finally, this energetic short GRB showed a bright optical and X-ray
afterglow, as well as a long-lived GeV emission, which could be
observed thanks to {\it Swift} and {\it Fermi} fast repoint
abilities. The temporal and spectral properties of this long-lasting emission 
could be studied over 9 energy decades and can be explained in at
least two ways : a combination of internal shock and forward shock
emission reproduces well the fluxes observed although some fine tuning
is required, and a forward-shock only origin can reproduce the
observed spectra although it requires some theoretical extensions.
Such joint {\it Swift-Fermi} observations are very
promising for understanding GRB afterglow physics.

\end{document}